\documentclass[article,twocolumn,preprintnumbers,superscriptaddress,nofootinbib]{revtex4-2}

\usepackage{enumerate}
\usepackage{mathrsfs,amsmath,amssymb}
\usepackage{natbib}
\usepackage{graphicx}
\usepackage{cancel}
\usepackage[normalem]{ulem}
\usepackage[pdftex,bookmarks,linktocpage,pdfpagelabels,plainpages=false,hyperfigures,linkcolor=blue,citecolor=blue]{hyperref} 
\hypersetup{colorlinks=true}
\usepackage{orcidlink}
\usepackage{pdfpages} 
\usepackage{pgffor} 

\makeatletter
\AtBeginDocument{\let\LS@rot\@undefined}
\makeatother

\def\supplementfilename{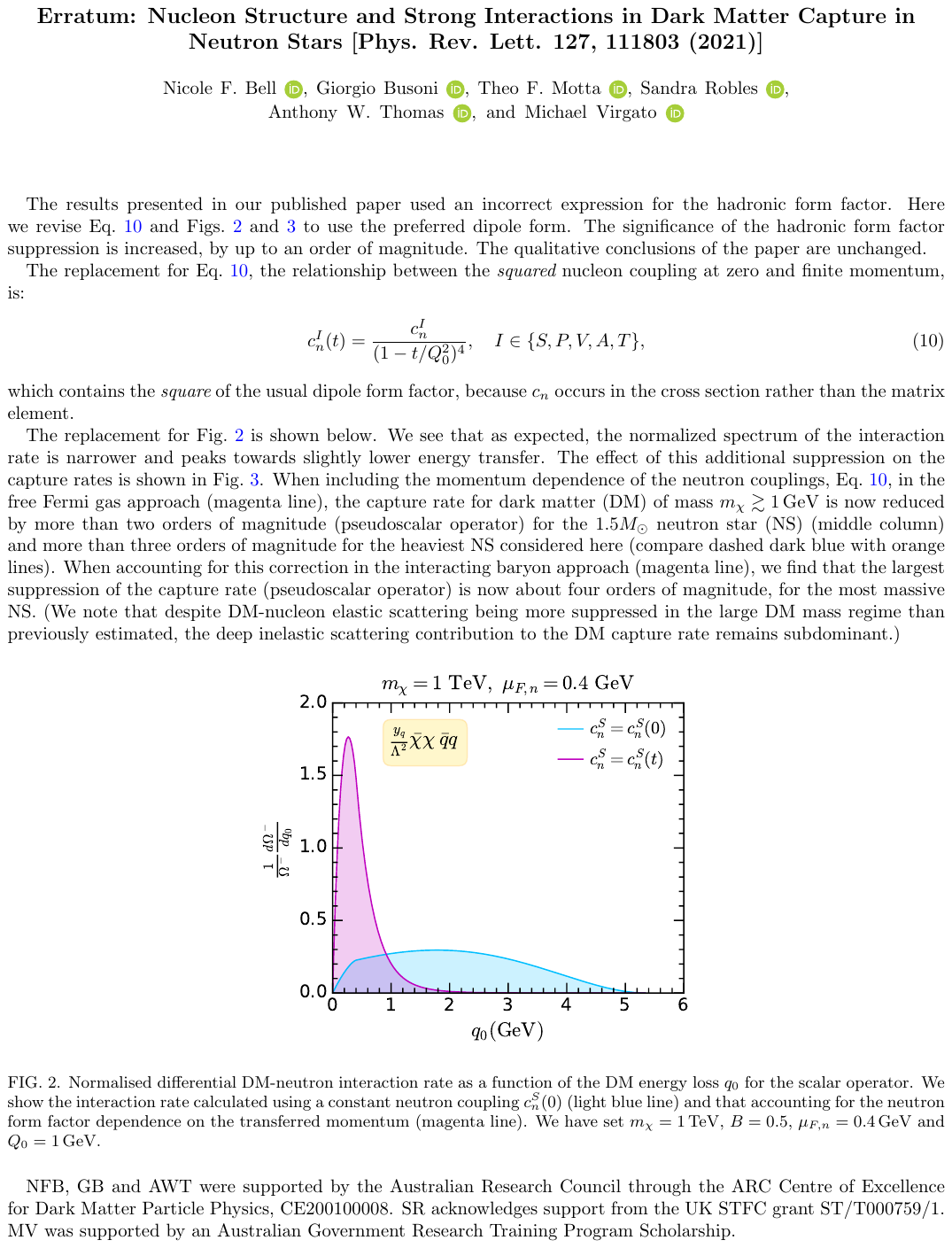}

\pdfximage{\supplementfilename}
\def\numbersupplementpages{\the\pdflastximagepages}

\newcommand{\GeV}{{\rm \,GeV}}
\newcommand{\TeV}{{\rm \,TeV}}

\newcommand{\cm}{\rm cm}

\newcommand{\Msun}{M_\odot}
\newcommand{\Mstar}{M_\star}
\newcommand{\Rstar}{R_\star}
\newcommand{\vstar}{v_\star}

\newcommand{\muFn}{\mu_{F,n}}

\newcommand{\fFD}{f_{\rm FD}}

\newcommand{\Msq}{|\overline{M}|^2}
\newcommand{\mneff}{m_n^{\rm eff}}

\begin{document}
\bibliographystyle{apsrev4-1}

\hfill \preprint{ADP-20-35/T1145}

\title{Nucleon Structure and Strong Interactions in Dark Matter Capture in Neutron Stars}

\author{Nicole F.~Bell~\orcidlink{0000-0002-5805-9828}}
\email{n.bell@unimelb.edu.au}
\affiliation{ARC Centre of Excellence for Dark Matter Particle Physics, School of Physics, The University of Melbourne, Victoria 3010, Australia}

\author{Giorgio Busoni~\orcidlink{0000-0002-8527-0768}}
\email{giorgio.busoni@mpi-hd.mpg.de}
\affiliation{Max-Planck-Institut fur Kernphysik, Saupfercheckweg 1, 69117 Heidelberg, Germany}

\author{Theo F. Motta~\orcidlink{0000-0003-3443-2496}}
\email{theo.ferrazmotta@adelaide.edu.au}
\affiliation{CSSM and ARC Centre of Excellence for Dark Matter Particle Physics, \\
Department of Physics, University of Adelaide, SA 5005 Australia}   
   
\author{Sandra Robles~\orcidlink{0000-0002-6046-8217}}
\email{sandra.robles@unimelb.edu.au}
\affiliation{ARC Centre of Excellence for Dark Matter Particle Physics, School of Physics, The University of Melbourne, Victoria 3010, Australia}

\author{\\ Anthony W. Thomas~\orcidlink{0000-0003-0026-499X}}
\email{anthony.thomas@adelaide.edu.au}
\affiliation{CSSM and ARC Centre of Excellence for Dark Matter Particle Physics, \\
Department of Physics, University of Adelaide, SA 5005 Australia}

\author{Michael Virgato~\orcidlink{0000-0002-8396-0896}}
\email{mvirgato@student.unimelb.edu.au}
\affiliation{ARC Centre of Excellence for Dark Matter Particle Physics, School of Physics, The University of Melbourne, Victoria 3010, Australia}

\begin{abstract}

We outline two important effects that are missing from most evaluations of the dark matter capture rate in neutron stars. As dark matter scattering with nucleons in the star involves large momentum transfer, nucleon structure must be taken into account via a momentum dependence of the hadronic form factors.  In addition, due to the high density of neutron star matter, we should account for nucleon interactions rather than modeling the nucleons as an ideal Fermi gas. Properly incorporating these effects is found to suppress the dark matter capture rate by up to three orders of magnitude for the heaviest stars.

\end{abstract}

\maketitle

\section{Introduction}
\label{sec:intro}

The scattering of cosmological dark matter (DM) particles with stars has long been used as a tool in the quest to uncover the particle nature of DM. For a wide range of DM masses, collisions between ambient DM and the constituents of a star would result in sufficient energy loss for the DM to become gravitationally bound to the star, with important observational consequences.  This provides a sensitive probe of dark matter scattering cross sections in a way that is highly complementary to terrestrial DM direct detection experiments.  While much attention has been focused on capture in the Sun~\citep{Gould:1987ju,Gould:1987ir,Jungman:1995df,Busoni:2013kaa,Garani:2017jcj,Busoni:2017mhe}, capture in neutron stars (NSs)~\citep{Goldman:1989nd,Kouvaris:2007ay,Kouvaris:2010vv,deLavallaz:2010wp,McDermott:2011jp,Bell:2013xk,Bramante:2013nma,Bramante:2017xlb,Baryakhtar:2017dbj,Raj:2017wrv,Bell:2018pkk,Garani:2018kkd,Camargo:2019wou,Bell:2019pyc,Acevedo:2019agu,Joglekar:2019vzy,Joglekar:2020liw,Bell:2020jou,Ilie:2020vec,Dasgupta:2020dik,Bell:2020lmm} has a similar long history.

The figure of merit for DM capture in NSs is the cross section for which the capture probability is of order 1.  Because of the enormous NS target mass and density, this extreme condition is met when the DM-neutron scattering cross section is $\sigma \sim 10^{-45} \cm^2$.  This is comparable to the sensitivity of DM-nucleon recoil experiments for those interactions for which direct detection is most sensitive, namely, unsuppressed spin-independent scattering of GeV scale DM. It is orders of magnitude more sensitive 
for high or low mass DM, spin-dependent interactions, or cross sections that are either velocity or momentum suppressed.  In many cases, this translates to a reach well below the so-called neutrino floor, beyond which neutrino scattering presents an irreducible background to direct detection experiments.

In recent years, there has been renewed interest in DM capture in NSs because of a number of key developments:  (i) the realization that DM capture, and subsequent annihilation, may lead to appreciable NS heating within reach of future telescopes in the near-infrared~\cite{Baryakhtar:2017dbj};  (ii) capture of non-annihilating DM, such as asymmetric DM, may trigger black hole formation~\cite{Kouvaris:2010jy,McDermott:2011jp,Kouvaris:2011fi,Bell:2013xk,Garani:2018kkd,Dasgupta:2020mqg}; (iii) improved understanding of NSs through a variety of observational data, including gravitational waves from NS mergers~\cite{TheLIGOScientific:2017qsa,Abbott:2018wiz,Monitor:2017mdv,Radice:2018ozg}.  However, up until recently, the treatment of DM capture in NSs has largely been adapted from that for capture in the Sun, without fully accounting for the extreme physics of a NS environment.

Due to the great promise of NS techniques, it is imperative to develop more accurate evaluations of the capture rate. To that end, recent calculations have included a fully relativistic scattering treatment~\cite{Joglekar:2019vzy,Joglekar:2020liw,Bell:2020jou,Bell:2020lmm}, gravitational focusing~\cite{Bell:2020jou,Bell:2020lmm}, Pauli blocking~\cite{Garani:2018kkd,Joglekar:2019vzy,Joglekar:2020liw,Bell:2020jou,Bell:2020lmm}, the opacity of the star~\citep{Bell:2020jou} and multiple-scattering effects ~\cite{Bramante:2017xlb,Joglekar:2019vzy,Joglekar:2020liw,Ilie:2020vec,Bell:2020jou,Bell:2020lmm}. In addition, one should properly incorporate the NS internal structure by consistently calculating the radial profiles of the equation of state (EoS) dependent parameters~\citep{Bell:2020jou,Garani:2018kkd,Bell:2020lmm} and the general relativistic corrections~\citep{Bell:2020jou,Bell:2020lmm}, by solving the Tolman-Oppenheimer-Volkoff (TOV) equations~\cite{Tolman:1939jz,Oppenheimer:1939ne}. However, despite these improvements, the current state-of-the-art calculations still miss important physical effects.

In this Letter, we address two important features that are intrinsic to the physics of neutron stars. In all prior treatments, the nucleon form factors that relate DM-nucleon couplings to the underlying DM-quark couplings have been evaluated at zero momentum transfer. While this is a valid assumption for non-relativistic scattering in direct detection experiments, it is a very poor approximation for the scattering of quasi-relativistic DM in a NS. Moreover, the nucleon targets are typically treated as a free Fermi gas, neglecting the fact that there are strong many body forces at play. 

We incorporate these effects for the first time, through (i) the use of momentum dependent form factors in the scattering matrix elements and (ii) effective masses  to account for strongly interacting nucleons.

\section{Capture of DM in neutron stars}

Neutron stars are primarily composed of degenerate neutrons. 
The simplest approach to calculate the DM capture rate, accounting for Pauli blocking, NS internal structure and general relativistic (GR) corrections, is to assume that DM scatters off a Fermi sea of neutrons, neglecting baryon interactions. Assuming that a single collision is sufficient for a DM particle to become gravitationally bound, the capture rate, $C$, is~\cite{Bell:2020jou}
\begin{widetext}
\begin{eqnarray}
C &=& \frac{4\pi}{\vstar} \frac{\rho_\chi}{m_\chi} {\rm Erf }\left(\sqrt{\frac{3}{2}}\frac{\vstar}{v_d}\right)\int_0^{\Rstar}  r^2 \frac{\sqrt{1-B(r)}}{B(r)} \Omega^{-}(r)  \, dr, \label{eq:captureM2} 
\\
\Omega^{-}(r) &=& \frac{\zeta(r)}{32\pi^3}\int dt dE_n ds  \frac{|\overline{M}(s,t)|^2}{s^2-(m_n^2-m_\chi^2)^2} \frac{E_n}{m_\chi}\sqrt{\frac{B(r)}{1-B(r)}}
\frac{s}{\gamma(s)}
\fFD(E_n,r)[1-\fFD(E_n^{'},r)],\label{eq:intrateideal}
\end{eqnarray}
\end{widetext}
where $r$ is the radial variable, $\Omega^{-}$ is the interaction rate, $B(r)$ is the time component of the Schwarzchild metric, $\rho_\chi$ is the local DM density, $\vstar$ is the NS velocity, $v_d$ is the DM velocity dispersion, $\Msq$ is the squared matrix element parametrized in terms of the Mandelstam variables $s$ and $t$, 
$E_n$ and $E_n^{'}$ are the initial and final neutron energies respectively, 
the Fermi Dirac distribution $\fFD$  depends on $r$ through the neutron chemical potential and
\begin{eqnarray}
    \gamma(s) &=& \sqrt{(s-m_n^2-m_\chi^2)^2-4m_n^2m_\chi^2},\\
       \zeta(r)&=&\frac{n_{n}(r)}{n_{free}(r)}, \label{eq:zeta}
\end{eqnarray}
where $n_{free}$ is the neutron number density in the ideal Fermi gas approximation. 
The integration intervals for $s$, $t$ and $E_n$ are given in Ref.~\cite{Bell:2020jou}. 

Evaluating Eqs.~\ref{eq:captureM2} and \ref{eq:intrateideal} requires the assumption of an equation of state (EoS) to determine
 realistic radial profiles for the neutron number density $n_n(r)$, chemical potential  and $B(r)$, by solving the general relativistic version of the equations of hydrostatic equilibrium, the TOV equations. 
The $\zeta(r)$ correction factor of Eq.~\ref{eq:zeta} was first introduced in Ref.~\cite{Garani:2018kkd} in order to retain the free Fermi gas approximation while using an EoS that accounts for nucleon interactions.  Most earlier calculations in the literature neglect this effect entirely, as they do not adopt an EoS and instead use average quantities.

\section{Nucleon Interactions}

At the extremely high densities found in neutron stars, particularly in the core, the ideal Fermi gas approach is no longer a good approximation, since nucleons undergo strong interactions. In equations of state of nucleon-rich dense matter, these interactions are often described in terms of effective Lagrangians such as Skyrme forces~\cite{Skyrme:1959zz,Vautherin:1971aw} and relativistic mean field models~\cite{Serot:1984ey,Serot:1997xg}. 
In the presence of a Lorentz scalar mean field, baryons in general, and nucleons in particular, develop an effective mass, $m^{\rm eff}$, different from the rest mass in vacuum, which must be used to consistently express the energy spectrum of interacting nucleons. Properly incorporating this effective mass in the evaluation of the capture rate is a superior approach to the use of the $\zeta(r)$ correction factor of Eq.~\ref{eq:zeta}.

To deal with the full range of NS masses, up to the maximum observed, a relativistic treatment is necessary. The constraints of $\beta$-equilibrium suggest that hyperons will appear at the high densities associated with stars above 1.6 $\Msun$. We use the EoS corresponding to the quark-meson coupling (QMC) model~\cite{Guichon:2018uew}, as presented in Ref.~\cite{Motta:2019tjc}. This model first suggested that one could have stars with masses of order 2 $\Msun$, even when hyperons appeared~\cite{RikovskaStone:2006ta}. This is a consequence of the repulsive three-body forces which arise naturally in that model~\cite{Guichon:2004xg} because of the self-consistent adjustment of the internal quark structure of the bound nucleons to the strong mean scalar field generated in the dense medium. 
In this model the energy of a neutron with momentum $\vec{p}_n$ is
\begin{equation}
 E_n(p_n)= \sqrt{p^2_n +\left[\mneff(n_b)\right]^2} +U_n(n_b), 
 \label{Eq:NeutronE}
\end{equation}
where $n_b$ is the baryon number density and $U_n$ is the Lorentz vector potential felt by the single neutron. Note that the kinetic term contains the effect of the Lorentz scalar interactions and the total energy resembles that of free particles~\cite{Reddy:1997yr}. 
 
The calculation of the DM-nucleon interaction rate is then similar to that for an ideal Fermi gas, but using $\mneff$ instead of the rest mass in Eq.~\ref{eq:intrateideal} and accounting for the available single neutron spectrum in the $\fFD$ distribution with the kinetic part of Eq.~\ref{Eq:NeutronE}. 
The interaction rate becomes

\begin{widetext}
\begin{equation}
\Omega^{-}(r) = \frac{1}{32\pi^3}\int dt dE_n ds  \frac{|\overline{M}(s,t,\mneff)|^2}{s^2-[(\mneff)^2-m_\chi^2]^2}
\frac{E_n}{m_\chi}\sqrt{\frac{B(r)}{1-B(r)}}\frac{s}{\gamma(s,\mneff)}\fFD(E_n,r)[1-\fFD(E_n^{'},r)].
\label{eq:intrate}
\end{equation}
\end{widetext}

\begin{figure}[t]
    \centering
    \includegraphics[width=8.5cm]{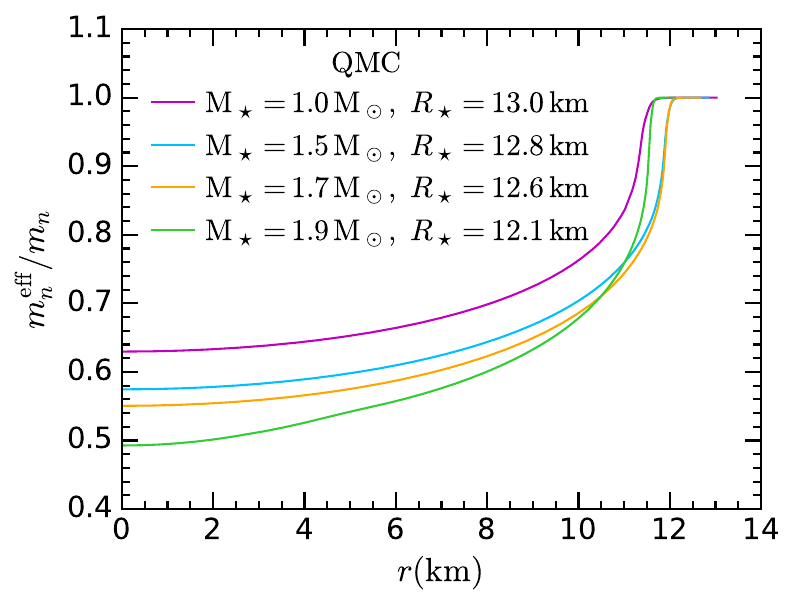}      
    \caption{Radial profile of the ratio of the neutron effective mass to the mass of isolated neutrons for different NS configurations of the QMC EoS, motivated by observations~\cite{Ozel:2016oaf,Antoniadis:2016hxz}.}
    \label{fig:mneff}
\end{figure}

The DM-neutron scattering rate, and the kinematically allowed phase space, now depend on the  radius 
through the neutron effective mass (which depends on the baryon number density). 
In Fig.~\ref{fig:mneff}, we show radial profiles of $\mneff$ for neutron stars of different masses with a QMC EoS. 
Note that the effective masses decrease with increasing density, towards the NS centre, and with heavier NSs.

\section{Nucleon Form Factors}

\begin{figure}[t] 
\centering
\includegraphics[width=8.5cm]{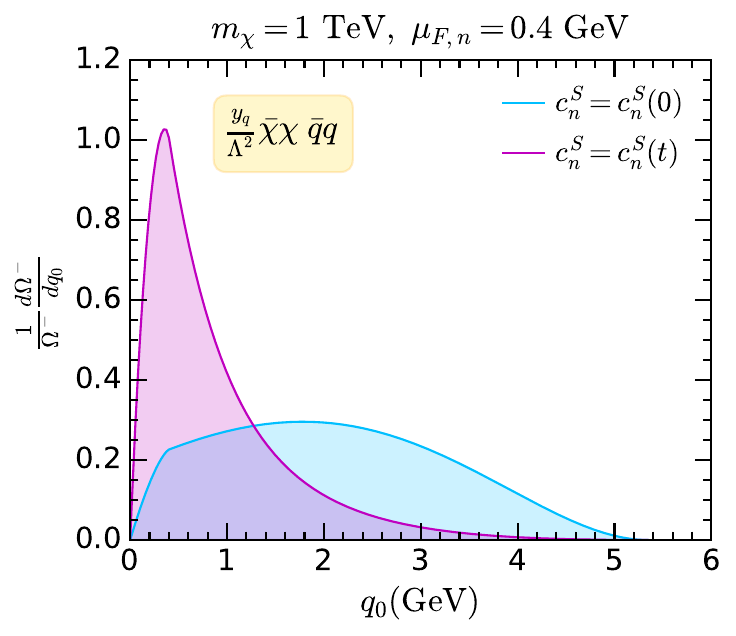}
\caption{Normalised differential DM-neutron interaction rate as a function of the DM energy loss $q_0$ for the scalar operator. 
We show the interaction rate calculated using a constant neutron coupling $c_n^S(0)$ (light blue line) and that accounting for the neutron form factor dependence  on the transferred momentum (magenta line). 
We have set $m_\chi=1\TeV$, $B=0.5$, $\muFn=0.4\GeV$ and $Q_0 =1\GeV$.}
\label{fig:intrateqtr1}
\end{figure}

When the momentum transfer in the DM-nucleon scattering process is sufficiently large, nucleons cannot be treated as point particles. This important observation has been overlooked in all existing NS capture calculations. Given the large DM velocity induced upon infall to the NS, 
it is necessary to account for the momentum dependence of the nucleon couplings when evaluating the scattering rates.\footnote{Note, however, that the momentum transfer is not large enough for deep inelastic scattering to make a significant contribution to the total cross section.}

To characterise these scattering rates, without loss of generality, we shall adopt an Effective Field Theory (EFT) framework to parametrize the coupling of DM to quarks and gluons.  For fermionic DM, the lowest order operators arise at mass dimension 6 and have the form $(\overline{\chi}\Gamma\chi)( \overline{q}\Gamma q)$, where $\Gamma$ represents the Lorentz structure of the operator. For scalar ($S$) and pseudoscalar ($P$) interactions, the operator coefficients are conventionally taken to scale as $y_q/\Lambda^2$, where  $y_q$ are the quark Yukawa couplings and $\Lambda$ is the cutoff scale of the effective theory. For vector ($V$), axial ($A$) and tensor ($T$) interactions, the operator coefficients are independent of the Yukawa couplings. 
The DM-quark couplings induce nucleon level interactions with protons and neutrons~\cite{DelNobile:2013sia}. 
The latter are usually evaluated at zero momentum transfer, a limit which is valid for direct detection experiments. The squared effective neutron couplings in this limit are
\begin{align}
c_n^S(\mneff) &= \frac{2 (\mneff)^2}{\Lambda^4 v^2 }\left[\sum_{q=u,d,s}f_{T_q}^{(n)}+\frac{2}{9}f_{T_G}^{(n)}\right]^2,\label{eq:scalarcoup}\\
c_n^P(\mneff) &= \frac{2 (\mneff)^2}{\Lambda^4 v^2}\left[\sum_{q=u,d,s}\left(1-3\frac{\overline{m}}{m_q}\right)\Delta_q^{(n)}\right]^2,\label{eq:pseudoscalarcoup}\\
c_n^V &= \frac{9}{\Lambda^4}, \qquad
c_n^A = \frac{1}{\Lambda^4}\left[\sum_{q=u,d,s}\Delta_q^{(n)}\right]^2, 
\end{align}
where $v=246$ GeV is the electroweak vacuum expectation value, $\overline{m}\equiv(1/m_u+1/m_d+1/m_s)^{-1}$ and 
 $f_{T_q}^{(n)}, f_{T_G}^{(n)}=1-\sum_{q=u,d,s} f_{T_q}^{(n)}$ and $\Delta_q^{(n)}$ are the hadronic matrix elements, determined via experiment or lattice QCD simulations.

\begin{figure*}[th]
    \centering
    \includegraphics[width=\textwidth]{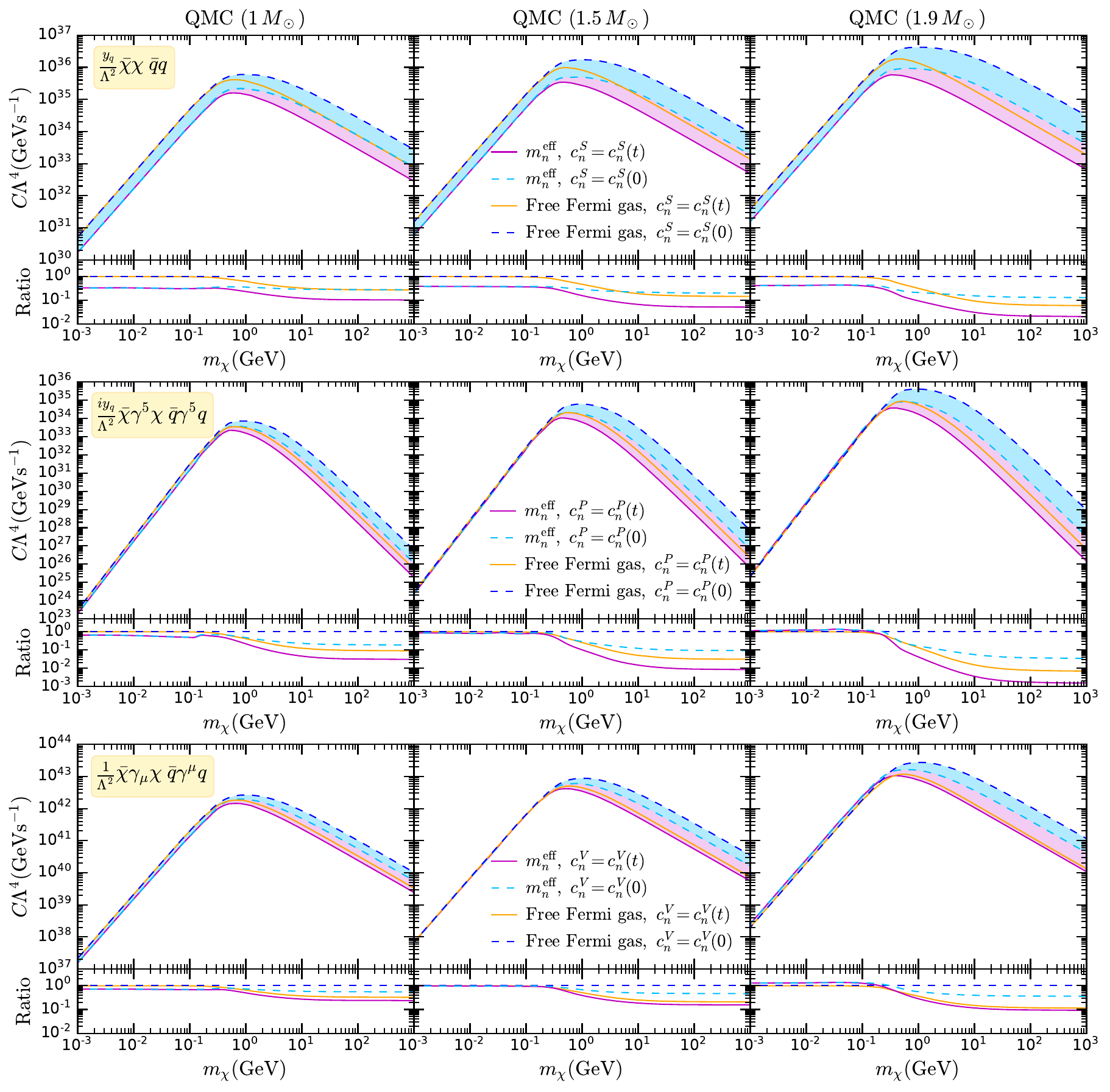}
    \caption{Capture rate in the optically thin limit for scalar (top), pseudoscalar (middle) and vector (bottom) operators as a function of the DM mass $m_\chi$, 
    using the free Fermi gas approach with constant neutron couplings (dashed blue line) and 
    momentum dependent couplings
    (orange line), and the interacting neutron approach 
    for constant nucleon couplings (dashed light blue line) and momentum dependent couplings 
    (magenta line), for NSs with a QCM EoS and $\Mstar=1\Msun$ (left),  $\Mstar=1.5\Msun$ (middle) and $\Mstar=1.9\Msun$ (right). In the lower panels, we show the ratios of the capture rates with respect to that for the free Fermi gas calculation with constant hadronic form factors. }
    \label{fig:Cmdm}
\end{figure*}

In contrast to direct detection, in neutron stars the transferred momentum could be of order $10\GeV$, depending on the NS and DM masses~\cite{Bell:2020jou}, and hence the 
momentum dependence of the nucleon couplings~\cite{Thomas:2001kw} should be included. The squared neutron form factors read
\begin{align}
c_n^I(t) &= \frac{c_n^I}{(1-t/Q_0^2)^2},\quad I\in\{S,P,V,A,T\},\label{eq:tdep}
\end{align}
where we have introduced the dependence on the transferred momentum through the Mandelstam variable $t$, and $Q_0$ is an energy scale 
that depends on the specific hadronic form factor. 
For simplicity, we take $Q_0=1\GeV$ for all operators, a conservative choice.

To illustrate the importance of correctly including this effect,  we show in Fig.~\ref{fig:intrateqtr1} the normalised differential interaction rate for the scalar operator, with (magenta) and without (light blue) accounting for the momentum dependence. Here we set the neutron mass equal to the rest mass. 
We see that the momentum dependence of Eq.~\ref{eq:tdep} suppresses the interaction rate when the energy transfer is large, shifting the normalized spectrum towards lower energy transfer.


\section{Results} 
\label{sec:results}

In Fig.~\ref{fig:Cmdm} we illustrate the impact of effective masses, and momentum dependent couplings, on the DM capture rate in NSs of mass  $1.0M_\odot$, $1.5M_\odot$ and $1.9M_\odot$, assuming the QMC EoS.
Results are shown for three representative choices of 
EFT operators, namely scalar, pseudoscalar and vector; the exact  expressions for their scattering amplitudes can be found in Table~2 of Ref.~\cite{Bell:2020jou}. 
Results are shown for the free Fermi gas approximation with neutron form factors at zero momentum transfer (dashed dark blue) and  momentum dependent form factors (orange), and for the interacting neutron approach, characterised by $\mneff$,  with (magenta) and without (dashed light blue) momentum dependent form factors. 
From these figures, we can conclude the following: 

\begin{enumerate}
\item   
The introduction of momentum-dependent form factors does not affect the capture rate when the DM mass is below $\sim0.2\GeV$, since the transferred momentum is small. (Compare orange line  with dashed dark blue line, or magenta line with dashed light blue line for any operator or NS mass.)

\item
The momentum dependence of the couplings strongly suppresses the interaction rate
at large DM mass, $m_\chi\gtrsim 1\GeV$. In fact, the form factors act as an effective cutoff on the kinematically available values of $t$ that are important for capture, removing large-$t$ contributions to the interaction rate (see Fig.~\ref{fig:intrateqtr1}). By including this correction in the free Fermi gas approximation, 
the capture rate is lowered by more than one order of magnitude (pseudoscalar operator) for the $1.5\Msun$ NS (middle column) and more than two orders of magnitude for the heaviest NS considered here (compare dashed dark blue with orange lines).

\item
For DM masses below $\sim0.2\GeV$ the capture rate is scarcely affected by the interacting baryon approach in the case of the  pseudoscalar (second row) and vector (bottom row) interactions. This low mass region corresponds to the Pauli blocking regime, where DM capture preferentially occurs close to the NS surface~\cite{Bell:2020jou} where neutrons are not  strongly degenerate and $\mneff\simeq m_n$ (see Fig.~\ref{fig:mneff}). For a scalar interaction (top panels) however, the use of $\mneff$  results in an overall rescaling (light blue shaded region) because the interaction rate scales with the neutron mass as $m_n^2$.   

\item 
Nucleon interactions significantly affect the capture when $m_\chi \gtrsim m_n$. This is true for all EFT operators. 
For the $1.5M_\odot$ NS, the capture rate is reduced by up to $\sim 1$ order of magnitude (see the light blue or magenta shaded regions, for constant or momentum dependent couplings, respectively).
This occur because the capture rate depends on the DM-neutron reduced mass, which approaches the neutron mass when $m_\chi\gg m_n$.  Unlike the capture of light DM, where Pauli blocking restricts the available neutron final states, DM of mass $m_\chi\gtrsim m_n$ can be captured deep in the star, where $\mneff<m_n$ (see Fig.~\ref{fig:mneff}). 
Thus, a lower $m_n$ induces a smaller reduced mass and hence a suppression of the capture rate.
 
\item
The suppression of the capture rate due to the momentum dependence of the form factors (dashed dark blue line vs orange line) is larger than that due to neutron interactions (dashed dark blue line vs dashed light blue line).
When both effects are included (magenta line) the total reduction is lower than the product of the two individual effects. The overall suppression is largest for the pseudoscalar operator
and large NS mass, where the reduction reaches up to 3 orders of magnitude.
\end{enumerate}

Finally, we estimate the uncertainties in our results:\\
(i) Based on the known values of the hadronic matrix elements~\cite{Zanotti:2017bte,Alarcon:2017ivh}, a conservative choice that holds for all operators is to take  $Q_0\sim0.9\pm0.1\GeV$. 
Comparing the capture rates for $Q_0 = 0.9\GeV$ and $Q_0 = 1\GeV$, we find that the smaller value of $Q_0$ results in a larger suppression in the capture rate  by a factor of order 1.2 -- 1.4, depending on the operator. In this sense, our choice of $Q_0 = 1\GeV$ is conservative.\\
(ii) We have compared our results for the QMC EoS with those for BSk24~\cite{Goriely:2013,Pearson:2018tkr,Chamel:2009yx}  (a Skyrme type EoS) and found little difference. E.g., the absolute difference in the capture rates are within 2--10\% for a $1.5\Msun$ NS, and within 10--30\% for a $1 \Msun$ NS, depending on the operator.


\section{Conclusion} 
\label{sec:conclusion}

The capture of dark matter in neutron stars has the potential to provide a very sensitive probe of dark matter interactions with ordinary matter. However, all previous treatments of the DM capture rate have neglected two important effects that are inherent to the physics of neutron stars.

First, treating nucleon targets as an ideal Fermi gas is a poor approximation when the capture occurs in the degenerate stellar interior, where strong interactions are expected to take place.
In the interacting nucleon framework, the neutron mass is essentially replaced by an effective mass. This radially dependent single neutron mass suppresses the capture of DM in the mass range $m_\chi \gtrsim m_n$. This effect  is stronger in denser NSs, where the neutron degeneracy is higher. 

Second, neutrons cannot be treated as point-like particles when calculating the DM-nucleon scattering cross section. Unlike dark matter direct detection experiments, where the limit of zero momentum transfer is always valid, dark matter scattering in a NS requires that the momentum dependence of the nucleon form factors be retained. This is necessary because the dark matter is accelerated to quasi relativistic velocities upon infall to a NS, due to the large gravitational field, resulting in collisions with appreciable momentum transfer when $m_\chi \gtrsim m_n$.  The suppression of the capture rate is most pronounced for heavier NSs, for which the gravitational fields are stronger.

We have shown that the combination of these two effects can suppress the dark matter capture rate by up to 3 orders of magnitude, for DM-neutron scattering.
We note that these effects are also relevant for DM scattering from other hadronic NS constituents, such as protons or hyperons. In addition, these effects are expected to have an even greater impact on the very heavy DM mass regime. 
This is because the reduced energy transfer per collision   (as illustrated in Fig.~\ref{fig:intrateqtr1}) implies that multi-scattering will be relevant for lower DM mass, and that a larger number of collisions will be required to achieve capture.

\begin{acknowledgments}
NFB, SR and AWT were supported in part by the Australian Research Council and MV by the Commonwealth of Australia. We thank Filippo Anzuini and Andrew Melatos for discussions.
\end{acknowledgments}

\bibliography{NSBib}

\begin{thebibliography}{54}%
\makeatletter
\providecommand \@ifxundefined [1]{%
 \@ifx{#1\undefined}
}%
\providecommand \@ifnum [1]{%
 \ifnum #1\expandafter \@firstoftwo
 \else \expandafter \@secondoftwo
 \fi
}%
\providecommand \@ifx [1]{%
 \ifx #1\expandafter \@firstoftwo
 \else \expandafter \@secondoftwo
 \fi
}%
\providecommand \natexlab [1]{#1}%
\providecommand \enquote  [1]{``#1''}%
\providecommand \bibnamefont  [1]{#1}%
\providecommand \bibfnamefont [1]{#1}%
\providecommand \citenamefont [1]{#1}%
\providecommand \href@noop [0]{\@secondoftwo}%
\providecommand \href [0]{\begingroup \@sanitize@url \@href}%
\providecommand \@href[1]{\@@startlink{#1}\@@href}%
\providecommand \@@href[1]{\endgroup#1\@@endlink}%
\providecommand \@sanitize@url [0]{\catcode `\\12\catcode `\$12\catcode
  `\&12\catcode `\#12\catcode `\^12\catcode `\_12\catcode `\%12\relax}%
\providecommand \@@startlink[1]{}%
\providecommand \@@endlink[0]{}%
\providecommand \url  [0]{\begingroup\@sanitize@url \@url }%
\providecommand \@url [1]{\endgroup\@href {#1}{\urlprefix }}%
\providecommand \urlprefix  [0]{URL }%
\providecommand \Eprint [0]{\href }%
\providecommand \doibase [0]{http://dx.doi.org/}%
\providecommand \selectlanguage [0]{\@gobble}%
\providecommand \bibinfo  [0]{\@secondoftwo}%
\providecommand \bibfield  [0]{\@secondoftwo}%
\providecommand \translation [1]{[#1]}%
\providecommand \BibitemOpen [0]{}%
\providecommand \bibitemStop [0]{}%
\providecommand \bibitemNoStop [0]{.\EOS\space}%
\providecommand \EOS [0]{\spacefactor3000\relax}%
\providecommand \BibitemShut  [1]{\csname bibitem#1\endcsname}%
\let\auto@bib@innerbib\@empty
\bibitem [{\citenamefont {Gould}(1987{\natexlab{a}})}]{Gould:1987ju}%
  \BibitemOpen
  \bibfield  {author} {\bibinfo {author} {\bibfnamefont {A.}~\bibnamefont
  {Gould}},\ }\href {\doibase 10.1086/165652} {\bibfield  {journal} {\bibinfo
  {journal} {Astrophys. J.}\ }\textbf {\bibinfo {volume} {321}},\ \bibinfo
  {pages} {560} (\bibinfo {year} {1987}{\natexlab{a}})}\BibitemShut {NoStop}%
\bibitem [{\citenamefont {Gould}(1987{\natexlab{b}})}]{Gould:1987ir}%
  \BibitemOpen
  \bibfield  {author} {\bibinfo {author} {\bibfnamefont {A.}~\bibnamefont
  {Gould}},\ }\href {\doibase 10.1086/165653} {\bibfield  {journal} {\bibinfo
  {journal} {Astrophys. J.}\ }\textbf {\bibinfo {volume} {321}},\ \bibinfo
  {pages} {571} (\bibinfo {year} {1987}{\natexlab{b}})}\BibitemShut {NoStop}%
\bibitem [{\citenamefont {Jungman}\ \emph {et~al.}(1996)\citenamefont
  {Jungman}, \citenamefont {Kamionkowski},\ and\ \citenamefont
  {Griest}}]{Jungman:1995df}%
  \BibitemOpen
  \bibfield  {author} {\bibinfo {author} {\bibfnamefont {G.}~\bibnamefont
  {Jungman}}, \bibinfo {author} {\bibfnamefont {M.}~\bibnamefont
  {Kamionkowski}}, \ and\ \bibinfo {author} {\bibfnamefont {K.}~\bibnamefont
  {Griest}},\ }\href {\doibase 10.1016/0370-1573(95)00058-5} {\bibfield
  {journal} {\bibinfo  {journal} {Phys. Rept.}\ }\textbf {\bibinfo {volume}
  {267}},\ \bibinfo {pages} {195} (\bibinfo {year} {1996})},\ \Eprint
  {http://arxiv.org/abs/hep-ph/9506380} {arXiv:hep-ph/9506380 [hep-ph]}
  \BibitemShut {NoStop}%
\bibitem [{\citenamefont {Busoni}\ \emph {et~al.}(2013)\citenamefont {Busoni},
  \citenamefont {De~Simone},\ and\ \citenamefont {Huang}}]{Busoni:2013kaa}%
  \BibitemOpen
  \bibfield  {author} {\bibinfo {author} {\bibfnamefont {G.}~\bibnamefont
  {Busoni}}, \bibinfo {author} {\bibfnamefont {A.}~\bibnamefont {De~Simone}}, \
  and\ \bibinfo {author} {\bibfnamefont {W.-C.}\ \bibnamefont {Huang}},\ }\href
  {\doibase 10.1088/1475-7516/2013/07/010} {\bibfield  {journal} {\bibinfo
  {journal} {JCAP}\ }\textbf {\bibinfo {volume} {1307}},\ \bibinfo {pages}
  {010} (\bibinfo {year} {2013})},\ \Eprint {http://arxiv.org/abs/1305.1817}
  {arXiv:1305.1817 [hep-ph]} \BibitemShut {NoStop}%
\bibitem [{\citenamefont {Garani}\ and\ \citenamefont
  {Palomares-Ruiz}(2017)}]{Garani:2017jcj}%
  \BibitemOpen
  \bibfield  {author} {\bibinfo {author} {\bibfnamefont {R.}~\bibnamefont
  {Garani}}\ and\ \bibinfo {author} {\bibfnamefont {S.}~\bibnamefont
  {Palomares-Ruiz}},\ }\href {\doibase 10.1088/1475-7516/2017/05/007}
  {\bibfield  {journal} {\bibinfo  {journal} {JCAP}\ }\textbf {\bibinfo
  {volume} {1705}},\ \bibinfo {pages} {007} (\bibinfo {year} {2017})},\ \Eprint
  {http://arxiv.org/abs/1702.02768} {arXiv:1702.02768 [hep-ph]} \BibitemShut
  {NoStop}%
\bibitem [{\citenamefont {Busoni}\ \emph {et~al.}(2017)\citenamefont {Busoni},
  \citenamefont {De~Simone}, \citenamefont {Scott},\ and\ \citenamefont
  {Vincent}}]{Busoni:2017mhe}%
  \BibitemOpen
  \bibfield  {author} {\bibinfo {author} {\bibfnamefont {G.}~\bibnamefont
  {Busoni}}, \bibinfo {author} {\bibfnamefont {A.}~\bibnamefont {De~Simone}},
  \bibinfo {author} {\bibfnamefont {P.}~\bibnamefont {Scott}}, \ and\ \bibinfo
  {author} {\bibfnamefont {A.~C.}\ \bibnamefont {Vincent}},\ }\href {\doibase
  10.1088/1475-7516/2017/10/037} {\bibfield  {journal} {\bibinfo  {journal}
  {JCAP}\ }\textbf {\bibinfo {volume} {1710}},\ \bibinfo {pages} {037}
  (\bibinfo {year} {2017})},\ \Eprint {http://arxiv.org/abs/1703.07784}
  {arXiv:1703.07784 [hep-ph]} \BibitemShut {NoStop}%
\bibitem [{\citenamefont {Goldman}\ and\ \citenamefont
  {Nussinov}(1989)}]{Goldman:1989nd}%
  \BibitemOpen
  \bibfield  {author} {\bibinfo {author} {\bibfnamefont {I.}~\bibnamefont
  {Goldman}}\ and\ \bibinfo {author} {\bibfnamefont {S.}~\bibnamefont
  {Nussinov}},\ }\href {\doibase 10.1103/PhysRevD.40.3221} {\bibfield
  {journal} {\bibinfo  {journal} {Phys. Rev.}\ }\textbf {\bibinfo {volume}
  {D40}},\ \bibinfo {pages} {3221} (\bibinfo {year} {1989})}\BibitemShut
  {NoStop}%
\bibitem [{\citenamefont {Kouvaris}(2008)}]{Kouvaris:2007ay}%
  \BibitemOpen
  \bibfield  {author} {\bibinfo {author} {\bibfnamefont {C.}~\bibnamefont
  {Kouvaris}},\ }\href {\doibase 10.1103/PhysRevD.77.023006} {\bibfield
  {journal} {\bibinfo  {journal} {Phys. Rev. D}\ }\textbf {\bibinfo {volume}
  {77}},\ \bibinfo {pages} {023006} (\bibinfo {year} {2008})},\ \Eprint
  {http://arxiv.org/abs/0708.2362} {arXiv:0708.2362 [astro-ph]} \BibitemShut
  {NoStop}%
\bibitem [{\citenamefont {Kouvaris}\ and\ \citenamefont
  {Tinyakov}(2010)}]{Kouvaris:2010vv}%
  \BibitemOpen
  \bibfield  {author} {\bibinfo {author} {\bibfnamefont {C.}~\bibnamefont
  {Kouvaris}}\ and\ \bibinfo {author} {\bibfnamefont {P.}~\bibnamefont
  {Tinyakov}},\ }\href {\doibase 10.1103/PhysRevD.82.063531} {\bibfield
  {journal} {\bibinfo  {journal} {Phys. Rev. D}\ }\textbf {\bibinfo {volume}
  {82}},\ \bibinfo {pages} {063531} (\bibinfo {year} {2010})},\ \Eprint
  {http://arxiv.org/abs/1004.0586} {arXiv:1004.0586 [astro-ph.GA]} \BibitemShut
  {NoStop}%
\bibitem [{\citenamefont {de~Lavallaz}\ and\ \citenamefont
  {Fairbairn}(2010)}]{deLavallaz:2010wp}%
  \BibitemOpen
  \bibfield  {author} {\bibinfo {author} {\bibfnamefont {A.}~\bibnamefont
  {de~Lavallaz}}\ and\ \bibinfo {author} {\bibfnamefont {M.}~\bibnamefont
  {Fairbairn}},\ }\href {\doibase 10.1103/PhysRevD.81.123521} {\bibfield
  {journal} {\bibinfo  {journal} {Phys. Rev. D}\ }\textbf {\bibinfo {volume}
  {81}},\ \bibinfo {pages} {123521} (\bibinfo {year} {2010})},\ \Eprint
  {http://arxiv.org/abs/1004.0629} {arXiv:1004.0629 [astro-ph.GA]} \BibitemShut
  {NoStop}%
\bibitem [{\citenamefont {McDermott}\ \emph {et~al.}(2012)\citenamefont
  {McDermott}, \citenamefont {Yu},\ and\ \citenamefont
  {Zurek}}]{McDermott:2011jp}%
  \BibitemOpen
  \bibfield  {author} {\bibinfo {author} {\bibfnamefont {S.~D.}\ \bibnamefont
  {McDermott}}, \bibinfo {author} {\bibfnamefont {H.-B.}\ \bibnamefont {Yu}}, \
  and\ \bibinfo {author} {\bibfnamefont {K.~M.}\ \bibnamefont {Zurek}},\ }\href
  {\doibase 10.1103/PhysRevD.85.023519} {\bibfield  {journal} {\bibinfo
  {journal} {Phys. Rev.}\ }\textbf {\bibinfo {volume} {D85}},\ \bibinfo {pages}
  {023519} (\bibinfo {year} {2012})},\ \Eprint {http://arxiv.org/abs/1103.5472}
  {arXiv:1103.5472 [hep-ph]} \BibitemShut {NoStop}%
\bibitem [{\citenamefont {Bell}\ \emph {et~al.}(2013)\citenamefont {Bell},
  \citenamefont {Melatos},\ and\ \citenamefont {Petraki}}]{Bell:2013xk}%
  \BibitemOpen
  \bibfield  {author} {\bibinfo {author} {\bibfnamefont {N.~F.}\ \bibnamefont
  {Bell}}, \bibinfo {author} {\bibfnamefont {A.}~\bibnamefont {Melatos}}, \
  and\ \bibinfo {author} {\bibfnamefont {K.}~\bibnamefont {Petraki}},\ }\href
  {\doibase 10.1103/PhysRevD.87.123507} {\bibfield  {journal} {\bibinfo
  {journal} {Phys. Rev.}\ }\textbf {\bibinfo {volume} {D87}},\ \bibinfo {pages}
  {123507} (\bibinfo {year} {2013})},\ \Eprint {http://arxiv.org/abs/1301.6811}
  {arXiv:1301.6811 [hep-ph]} \BibitemShut {NoStop}%
\bibitem [{\citenamefont {Bramante}\ \emph {et~al.}(2014)\citenamefont
  {Bramante}, \citenamefont {Fukushima}, \citenamefont {Kumar},\ and\
  \citenamefont {Stopnitzky}}]{Bramante:2013nma}%
  \BibitemOpen
  \bibfield  {author} {\bibinfo {author} {\bibfnamefont {J.}~\bibnamefont
  {Bramante}}, \bibinfo {author} {\bibfnamefont {K.}~\bibnamefont {Fukushima}},
  \bibinfo {author} {\bibfnamefont {J.}~\bibnamefont {Kumar}}, \ and\ \bibinfo
  {author} {\bibfnamefont {E.}~\bibnamefont {Stopnitzky}},\ }\href {\doibase
  10.1103/PhysRevD.89.015010} {\bibfield  {journal} {\bibinfo  {journal} {Phys.
  Rev.}\ }\textbf {\bibinfo {volume} {D89}},\ \bibinfo {pages} {015010}
  (\bibinfo {year} {2014})},\ \Eprint {http://arxiv.org/abs/1310.3509}
  {arXiv:1310.3509 [hep-ph]} \BibitemShut {NoStop}%
\bibitem [{\citenamefont {Bramante}\ \emph {et~al.}(2017)\citenamefont
  {Bramante}, \citenamefont {Delgado},\ and\ \citenamefont
  {Martin}}]{Bramante:2017xlb}%
  \BibitemOpen
  \bibfield  {author} {\bibinfo {author} {\bibfnamefont {J.}~\bibnamefont
  {Bramante}}, \bibinfo {author} {\bibfnamefont {A.}~\bibnamefont {Delgado}}, \
  and\ \bibinfo {author} {\bibfnamefont {A.}~\bibnamefont {Martin}},\ }\href
  {\doibase 10.1103/PhysRevD.96.063002} {\bibfield  {journal} {\bibinfo
  {journal} {Phys. Rev.}\ }\textbf {\bibinfo {volume} {D96}},\ \bibinfo {pages}
  {063002} (\bibinfo {year} {2017})},\ \Eprint
  {http://arxiv.org/abs/1703.04043} {arXiv:1703.04043 [hep-ph]} \BibitemShut
  {NoStop}%
\bibitem [{\citenamefont {Baryakhtar}\ \emph {et~al.}(2017)\citenamefont
  {Baryakhtar}, \citenamefont {Bramante}, \citenamefont {Li}, \citenamefont
  {Linden},\ and\ \citenamefont {Raj}}]{Baryakhtar:2017dbj}%
  \BibitemOpen
  \bibfield  {author} {\bibinfo {author} {\bibfnamefont {M.}~\bibnamefont
  {Baryakhtar}}, \bibinfo {author} {\bibfnamefont {J.}~\bibnamefont
  {Bramante}}, \bibinfo {author} {\bibfnamefont {S.~W.}\ \bibnamefont {Li}},
  \bibinfo {author} {\bibfnamefont {T.}~\bibnamefont {Linden}}, \ and\ \bibinfo
  {author} {\bibfnamefont {N.}~\bibnamefont {Raj}},\ }\href {\doibase
  10.1103/PhysRevLett.119.131801} {\bibfield  {journal} {\bibinfo  {journal}
  {Phys. Rev. Lett.}\ }\textbf {\bibinfo {volume} {119}},\ \bibinfo {pages}
  {131801} (\bibinfo {year} {2017})},\ \Eprint
  {http://arxiv.org/abs/1704.01577} {arXiv:1704.01577 [hep-ph]} \BibitemShut
  {NoStop}%
\bibitem [{\citenamefont {Raj}\ \emph {et~al.}(2018)\citenamefont {Raj},
  \citenamefont {Tanedo},\ and\ \citenamefont {Yu}}]{Raj:2017wrv}%
  \BibitemOpen
  \bibfield  {author} {\bibinfo {author} {\bibfnamefont {N.}~\bibnamefont
  {Raj}}, \bibinfo {author} {\bibfnamefont {P.}~\bibnamefont {Tanedo}}, \ and\
  \bibinfo {author} {\bibfnamefont {H.-B.}\ \bibnamefont {Yu}},\ }\href
  {\doibase 10.1103/PhysRevD.97.043006} {\bibfield  {journal} {\bibinfo
  {journal} {Phys. Rev.}\ }\textbf {\bibinfo {volume} {D97}},\ \bibinfo {pages}
  {043006} (\bibinfo {year} {2018})},\ \Eprint
  {http://arxiv.org/abs/1707.09442} {arXiv:1707.09442 [hep-ph]} \BibitemShut
  {NoStop}%
\bibitem [{\citenamefont {Bell}\ \emph {et~al.}(2018)\citenamefont {Bell},
  \citenamefont {Busoni},\ and\ \citenamefont {Robles}}]{Bell:2018pkk}%
  \BibitemOpen
  \bibfield  {author} {\bibinfo {author} {\bibfnamefont {N.~F.}\ \bibnamefont
  {Bell}}, \bibinfo {author} {\bibfnamefont {G.}~\bibnamefont {Busoni}}, \ and\
  \bibinfo {author} {\bibfnamefont {S.}~\bibnamefont {Robles}},\ }\href
  {\doibase 10.1088/1475-7516/2018/09/018} {\bibfield  {journal} {\bibinfo
  {journal} {JCAP}\ }\textbf {\bibinfo {volume} {1809}},\ \bibinfo {pages}
  {018} (\bibinfo {year} {2018})},\ \Eprint {http://arxiv.org/abs/1807.02840}
  {arXiv:1807.02840 [hep-ph]} \BibitemShut {NoStop}%
\bibitem [{\citenamefont {Garani}\ \emph {et~al.}(2019)\citenamefont {Garani},
  \citenamefont {Genolini},\ and\ \citenamefont {Hambye}}]{Garani:2018kkd}%
  \BibitemOpen
  \bibfield  {author} {\bibinfo {author} {\bibfnamefont {R.}~\bibnamefont
  {Garani}}, \bibinfo {author} {\bibfnamefont {Y.}~\bibnamefont {Genolini}}, \
  and\ \bibinfo {author} {\bibfnamefont {T.}~\bibnamefont {Hambye}},\ }\href
  {\doibase 10.1088/1475-7516/2019/05/035} {\bibfield  {journal} {\bibinfo
  {journal} {JCAP}\ }\textbf {\bibinfo {volume} {05}},\ \bibinfo {pages} {035}
  (\bibinfo {year} {2019})},\ \Eprint {http://arxiv.org/abs/1812.08773}
  {arXiv:1812.08773 [hep-ph]} \BibitemShut {NoStop}%
\bibitem [{\citenamefont {Camargo}\ \emph {et~al.}(2019)\citenamefont
  {Camargo}, \citenamefont {Queiroz},\ and\ \citenamefont
  {Sturani}}]{Camargo:2019wou}%
  \BibitemOpen
  \bibfield  {author} {\bibinfo {author} {\bibfnamefont {D.~A.}\ \bibnamefont
  {Camargo}}, \bibinfo {author} {\bibfnamefont {F.~S.}\ \bibnamefont
  {Queiroz}}, \ and\ \bibinfo {author} {\bibfnamefont {R.}~\bibnamefont
  {Sturani}},\ }\href {\doibase 10.1088/1475-7516/2019/09/051} {\bibfield
  {journal} {\bibinfo  {journal} {JCAP}\ }\textbf {\bibinfo {volume} {09}},\
  \bibinfo {pages} {051} (\bibinfo {year} {2019})},\ \Eprint
  {http://arxiv.org/abs/1901.05474} {arXiv:1901.05474 [hep-ph]} \BibitemShut
  {NoStop}%
\bibitem [{\citenamefont {Bell}\ \emph {et~al.}(2019)\citenamefont {Bell},
  \citenamefont {Busoni},\ and\ \citenamefont {Robles}}]{Bell:2019pyc}%
  \BibitemOpen
  \bibfield  {author} {\bibinfo {author} {\bibfnamefont {N.~F.}\ \bibnamefont
  {Bell}}, \bibinfo {author} {\bibfnamefont {G.}~\bibnamefont {Busoni}}, \ and\
  \bibinfo {author} {\bibfnamefont {S.}~\bibnamefont {Robles}},\ }\href
  {\doibase 10.1088/1475-7516/2019/06/054} {\bibfield  {journal} {\bibinfo
  {journal} {JCAP}\ }\textbf {\bibinfo {volume} {1906}},\ \bibinfo {pages}
  {054} (\bibinfo {year} {2019})},\ \Eprint {http://arxiv.org/abs/1904.09803}
  {arXiv:1904.09803 [hep-ph]} \BibitemShut {NoStop}%
\bibitem [{\citenamefont {Acevedo}\ \emph {et~al.}(2020)\citenamefont
  {Acevedo}, \citenamefont {Bramante}, \citenamefont {Leane},\ and\
  \citenamefont {Raj}}]{Acevedo:2019agu}%
  \BibitemOpen
  \bibfield  {author} {\bibinfo {author} {\bibfnamefont {J.~F.}\ \bibnamefont
  {Acevedo}}, \bibinfo {author} {\bibfnamefont {J.}~\bibnamefont {Bramante}},
  \bibinfo {author} {\bibfnamefont {R.~K.}\ \bibnamefont {Leane}}, \ and\
  \bibinfo {author} {\bibfnamefont {N.}~\bibnamefont {Raj}},\ }\href {\doibase
  10.1088/1475-7516/2020/03/038} {\bibfield  {journal} {\bibinfo  {journal}
  {JCAP}\ }\textbf {\bibinfo {volume} {03}},\ \bibinfo {pages} {038} (\bibinfo
  {year} {2020})},\ \Eprint {http://arxiv.org/abs/1911.06334} {arXiv:1911.06334
  [hep-ph]} \BibitemShut {NoStop}%
\bibitem [{\citenamefont {Joglekar}\ \emph
  {et~al.}(2020{\natexlab{a}})\citenamefont {Joglekar}, \citenamefont {Raj},
  \citenamefont {Tanedo},\ and\ \citenamefont {Yu}}]{Joglekar:2019vzy}%
  \BibitemOpen
  \bibfield  {author} {\bibinfo {author} {\bibfnamefont {A.}~\bibnamefont
  {Joglekar}}, \bibinfo {author} {\bibfnamefont {N.}~\bibnamefont {Raj}},
  \bibinfo {author} {\bibfnamefont {P.}~\bibnamefont {Tanedo}}, \ and\ \bibinfo
  {author} {\bibfnamefont {H.-B.}\ \bibnamefont {Yu}},\ }\href {\doibase
  10.1016/j.physletb.2020.135767} {\bibfield  {journal} {\bibinfo  {journal}
  {Phys. Lett.}\ }\textbf {\bibinfo {volume} {B}},\ \bibinfo {pages} {135767}
  (\bibinfo {year} {2020}{\natexlab{a}})},\ \Eprint
  {http://arxiv.org/abs/1911.13293} {arXiv:1911.13293 [hep-ph]} \BibitemShut
  {NoStop}%
\bibitem [{\citenamefont {Joglekar}\ \emph
  {et~al.}(2020{\natexlab{b}})\citenamefont {Joglekar}, \citenamefont {Raj},
  \citenamefont {Tanedo},\ and\ \citenamefont {Yu}}]{Joglekar:2020liw}%
  \BibitemOpen
  \bibfield  {author} {\bibinfo {author} {\bibfnamefont {A.}~\bibnamefont
  {Joglekar}}, \bibinfo {author} {\bibfnamefont {N.}~\bibnamefont {Raj}},
  \bibinfo {author} {\bibfnamefont {P.}~\bibnamefont {Tanedo}}, \ and\ \bibinfo
  {author} {\bibfnamefont {H.-B.}\ \bibnamefont {Yu}},\ }\href {\doibase
  10.1103/PhysRevD.102.123002} {\bibfield  {journal} {\bibinfo  {journal}
  {Phys. Rev. D}\ }\textbf {\bibinfo {volume} {102}},\ \bibinfo {pages}
  {123002} (\bibinfo {year} {2020}{\natexlab{b}})},\ \Eprint
  {http://arxiv.org/abs/2004.09539} {arXiv:2004.09539 [hep-ph]} \BibitemShut
  {NoStop}%
\bibitem [{\citenamefont {Bell}\ \emph {et~al.}(2020)\citenamefont {Bell},
  \citenamefont {Busoni}, \citenamefont {Robles},\ and\ \citenamefont
  {Virgato}}]{Bell:2020jou}%
  \BibitemOpen
  \bibfield  {author} {\bibinfo {author} {\bibfnamefont {N.~F.}\ \bibnamefont
  {Bell}}, \bibinfo {author} {\bibfnamefont {G.}~\bibnamefont {Busoni}},
  \bibinfo {author} {\bibfnamefont {S.}~\bibnamefont {Robles}}, \ and\ \bibinfo
  {author} {\bibfnamefont {M.}~\bibnamefont {Virgato}},\ }\href {\doibase
  10.1088/1475-7516/2020/09/028} {\bibfield  {journal} {\bibinfo  {journal}
  {JCAP}\ }\textbf {\bibinfo {volume} {09}},\ \bibinfo {pages} {028} (\bibinfo
  {year} {2020})},\ \Eprint {http://arxiv.org/abs/2004.14888} {arXiv:2004.14888
  [hep-ph]} \BibitemShut {NoStop}%
\bibitem [{\citenamefont {Ilie}\ \emph {et~al.}(2020)\citenamefont {Ilie},
  \citenamefont {Pilawa},\ and\ \citenamefont {Zhang}}]{Ilie:2020vec}%
  \BibitemOpen
  \bibfield  {author} {\bibinfo {author} {\bibfnamefont {C.}~\bibnamefont
  {Ilie}}, \bibinfo {author} {\bibfnamefont {J.}~\bibnamefont {Pilawa}}, \ and\
  \bibinfo {author} {\bibfnamefont {S.}~\bibnamefont {Zhang}},\ }\href
  {\doibase 10.1103/PhysRevD.102.048301} {\bibfield  {journal} {\bibinfo
  {journal} {Phys. Rev. D}\ }\textbf {\bibinfo {volume} {102}},\ \bibinfo
  {pages} {048301} (\bibinfo {year} {2020})},\ \Eprint
  {http://arxiv.org/abs/2005.05946} {arXiv:2005.05946 [astro-ph.CO]}
  \BibitemShut {NoStop}%
\bibitem [{\citenamefont {Dasgupta}\ \emph
  {et~al.}(2020{\natexlab{a}})\citenamefont {Dasgupta}, \citenamefont {Gupta},\
  and\ \citenamefont {Ray}}]{Dasgupta:2020dik}%
  \BibitemOpen
  \bibfield  {author} {\bibinfo {author} {\bibfnamefont {B.}~\bibnamefont
  {Dasgupta}}, \bibinfo {author} {\bibfnamefont {A.}~\bibnamefont {Gupta}}, \
  and\ \bibinfo {author} {\bibfnamefont {A.}~\bibnamefont {Ray}},\ }\href
  {\doibase 10.1088/1475-7516/2020/10/023} {\bibfield  {journal} {\bibinfo
  {journal} {JCAP}\ }\textbf {\bibinfo {volume} {10}},\ \bibinfo {pages} {023}
  (\bibinfo {year} {2020}{\natexlab{a}})},\ \Eprint
  {http://arxiv.org/abs/2006.10773} {arXiv:2006.10773 [hep-ph]} \BibitemShut
  {NoStop}%
\bibitem [{\citenamefont {Bell}\ \emph {et~al.}(2021)\citenamefont {Bell},
  \citenamefont {Busoni}, \citenamefont {Robles},\ and\ \citenamefont
  {Virgato}}]{Bell:2020lmm}%
  \BibitemOpen
  \bibfield  {author} {\bibinfo {author} {\bibfnamefont {N.~F.}\ \bibnamefont
  {Bell}}, \bibinfo {author} {\bibfnamefont {G.}~\bibnamefont {Busoni}},
  \bibinfo {author} {\bibfnamefont {S.}~\bibnamefont {Robles}}, \ and\ \bibinfo
  {author} {\bibfnamefont {M.}~\bibnamefont {Virgato}},\ }\href {\doibase
  10.1088/1475-7516/2021/03/086} {\bibfield  {journal} {\bibinfo  {journal}
  {JCAP}\ }\textbf {\bibinfo {volume} {03}},\ \bibinfo {pages} {086} (\bibinfo
  {year} {2021})},\ \Eprint {http://arxiv.org/abs/2010.13257} {arXiv:2010.13257
  [hep-ph]} \BibitemShut {NoStop}%
\bibitem [{\citenamefont {Kouvaris}\ and\ \citenamefont
  {Tinyakov}(2011{\natexlab{a}})}]{Kouvaris:2010jy}%
  \BibitemOpen
  \bibfield  {author} {\bibinfo {author} {\bibfnamefont {C.}~\bibnamefont
  {Kouvaris}}\ and\ \bibinfo {author} {\bibfnamefont {P.}~\bibnamefont
  {Tinyakov}},\ }\href {\doibase 10.1103/PhysRevD.83.083512} {\bibfield
  {journal} {\bibinfo  {journal} {Phys. Rev.}\ }\textbf {\bibinfo {volume}
  {D83}},\ \bibinfo {pages} {083512} (\bibinfo {year} {2011}{\natexlab{a}})},\
  \Eprint {http://arxiv.org/abs/1012.2039} {arXiv:1012.2039 [astro-ph.HE]}
  \BibitemShut {NoStop}%
\bibitem [{\citenamefont {Kouvaris}\ and\ \citenamefont
  {Tinyakov}(2011{\natexlab{b}})}]{Kouvaris:2011fi}%
  \BibitemOpen
  \bibfield  {author} {\bibinfo {author} {\bibfnamefont {C.}~\bibnamefont
  {Kouvaris}}\ and\ \bibinfo {author} {\bibfnamefont {P.}~\bibnamefont
  {Tinyakov}},\ }\href {\doibase 10.1103/PhysRevLett.107.091301} {\bibfield
  {journal} {\bibinfo  {journal} {Phys. Rev. Lett.}\ }\textbf {\bibinfo
  {volume} {107}},\ \bibinfo {pages} {091301} (\bibinfo {year}
  {2011}{\natexlab{b}})},\ \Eprint {http://arxiv.org/abs/1104.0382}
  {arXiv:1104.0382 [astro-ph.CO]} \BibitemShut {NoStop}%
\bibitem [{\citenamefont {Dasgupta}\ \emph
  {et~al.}(2020{\natexlab{b}})\citenamefont {Dasgupta}, \citenamefont {Laha},\
  and\ \citenamefont {Ray}}]{Dasgupta:2020mqg}%
  \BibitemOpen
  \bibfield  {author} {\bibinfo {author} {\bibfnamefont {B.}~\bibnamefont
  {Dasgupta}}, \bibinfo {author} {\bibfnamefont {R.}~\bibnamefont {Laha}}, \
  and\ \bibinfo {author} {\bibfnamefont {A.}~\bibnamefont {Ray}},\ }\href@noop
  {} {\  (\bibinfo {year} {2020}{\natexlab{b}})},\ \Eprint
  {http://arxiv.org/abs/2009.01825} {arXiv:2009.01825 [astro-ph.HE]}
  \BibitemShut {NoStop}%
\bibitem [{\citenamefont {Abbott}\ \emph
  {et~al.}(2017{\natexlab{a}})\citenamefont {Abbott} \emph
  {et~al.}}]{TheLIGOScientific:2017qsa}%
  \BibitemOpen
  \bibfield  {author} {\bibinfo {author} {\bibfnamefont {B.~P.}\ \bibnamefont
  {Abbott}} \emph {et~al.} (\bibinfo {collaboration} {LIGO Scientific,
  Virgo}),\ }\href {\doibase 10.1103/PhysRevLett.119.161101} {\bibfield
  {journal} {\bibinfo  {journal} {Phys. Rev. Lett.}\ }\textbf {\bibinfo
  {volume} {119}},\ \bibinfo {pages} {161101} (\bibinfo {year}
  {2017}{\natexlab{a}})},\ \Eprint {http://arxiv.org/abs/1710.05832}
  {arXiv:1710.05832 [gr-qc]} \BibitemShut {NoStop}%
\bibitem [{\citenamefont {Abbott}\ \emph {et~al.}(2019)\citenamefont {Abbott}
  \emph {et~al.}}]{Abbott:2018wiz}%
  \BibitemOpen
  \bibfield  {author} {\bibinfo {author} {\bibfnamefont {B.~P.}\ \bibnamefont
  {Abbott}} \emph {et~al.} (\bibinfo {collaboration} {LIGO Scientific,
  Virgo}),\ }\href {\doibase 10.1103/PhysRevX.9.011001} {\bibfield  {journal}
  {\bibinfo  {journal} {Phys. Rev.}\ }\textbf {\bibinfo {volume} {X9}},\
  \bibinfo {pages} {011001} (\bibinfo {year} {2019})},\ \Eprint
  {http://arxiv.org/abs/1805.11579} {arXiv:1805.11579 [gr-qc]} \BibitemShut
  {NoStop}%
\bibitem [{\citenamefont {Abbott}\ \emph
  {et~al.}(2017{\natexlab{b}})\citenamefont {Abbott} \emph
  {et~al.}}]{Monitor:2017mdv}%
  \BibitemOpen
  \bibfield  {author} {\bibinfo {author} {\bibfnamefont {B.~P.}\ \bibnamefont
  {Abbott}} \emph {et~al.} (\bibinfo {collaboration} {LIGO Scientific, Virgo,
  Fermi-GBM, INTEGRAL}),\ }\href {\doibase 10.3847/2041-8213/aa920c} {\bibfield
   {journal} {\bibinfo  {journal} {Astrophys. J.}\ }\textbf {\bibinfo {volume}
  {848}},\ \bibinfo {pages} {L13} (\bibinfo {year} {2017}{\natexlab{b}})},\
  \Eprint {http://arxiv.org/abs/1710.05834} {arXiv:1710.05834 [astro-ph.HE]}
  \BibitemShut {NoStop}%
\bibitem [{\citenamefont {Radice}\ and\ \citenamefont
  {Dai}(2019)}]{Radice:2018ozg}%
  \BibitemOpen
  \bibfield  {author} {\bibinfo {author} {\bibfnamefont {D.}~\bibnamefont
  {Radice}}\ and\ \bibinfo {author} {\bibfnamefont {L.}~\bibnamefont {Dai}},\
  }\href {\doibase 10.1140/epja/i2019-12716-4} {\bibfield  {journal} {\bibinfo
  {journal} {Eur. Phys. J.}\ }\textbf {\bibinfo {volume} {A55}},\ \bibinfo
  {pages} {50} (\bibinfo {year} {2019})},\ \Eprint
  {http://arxiv.org/abs/1810.12917} {arXiv:1810.12917 [astro-ph.HE]}
  \BibitemShut {NoStop}%
\bibitem [{\citenamefont {Tolman}(1939)}]{Tolman:1939jz}%
  \BibitemOpen
  \bibfield  {author} {\bibinfo {author} {\bibfnamefont {R.~C.}\ \bibnamefont
  {Tolman}},\ }\href {\doibase 10.1103/PhysRev.55.364} {\bibfield  {journal}
  {\bibinfo  {journal} {Phys. Rev.}\ }\textbf {\bibinfo {volume} {55}},\
  \bibinfo {pages} {364} (\bibinfo {year} {1939})}\BibitemShut {NoStop}%
\bibitem [{\citenamefont {Oppenheimer}\ and\ \citenamefont
  {Volkoff}(1939)}]{Oppenheimer:1939ne}%
  \BibitemOpen
  \bibfield  {author} {\bibinfo {author} {\bibfnamefont {J.~R.}\ \bibnamefont
  {Oppenheimer}}\ and\ \bibinfo {author} {\bibfnamefont {G.~M.}\ \bibnamefont
  {Volkoff}},\ }\href {\doibase 10.1103/PhysRev.55.374} {\bibfield  {journal}
  {\bibinfo  {journal} {Phys. Rev.}\ }\textbf {\bibinfo {volume} {55}},\
  \bibinfo {pages} {374} (\bibinfo {year} {1939})}\BibitemShut {NoStop}%
\bibitem [{\citenamefont {Skyrme}(1958)}]{Skyrme:1959zz}%
  \BibitemOpen
  \bibfield  {author} {\bibinfo {author} {\bibfnamefont {T.}~\bibnamefont
  {Skyrme}},\ }\href {\doibase 10.1016/0029-5582(58)90345-6} {\bibfield
  {journal} {\bibinfo  {journal} {Nucl. Phys.}\ }\textbf {\bibinfo {volume}
  {9}},\ \bibinfo {pages} {615} (\bibinfo {year} {1958})}\BibitemShut {NoStop}%
\bibitem [{\citenamefont {Vautherin}\ and\ \citenamefont
  {Brink}(1972)}]{Vautherin:1971aw}%
  \BibitemOpen
  \bibfield  {author} {\bibinfo {author} {\bibfnamefont {D.}~\bibnamefont
  {Vautherin}}\ and\ \bibinfo {author} {\bibfnamefont {D.}~\bibnamefont
  {Brink}},\ }\href {\doibase 10.1103/PhysRevC.5.626} {\bibfield  {journal}
  {\bibinfo  {journal} {Phys. Rev. C}\ }\textbf {\bibinfo {volume} {5}},\
  \bibinfo {pages} {626} (\bibinfo {year} {1972})}\BibitemShut {NoStop}%
\bibitem [{\citenamefont {Serot}\ and\ \citenamefont
  {Walecka}(1986)}]{Serot:1984ey}%
  \BibitemOpen
  \bibfield  {author} {\bibinfo {author} {\bibfnamefont {B.~D.}\ \bibnamefont
  {Serot}}\ and\ \bibinfo {author} {\bibfnamefont {J.~D.}\ \bibnamefont
  {Walecka}},\ }\href@noop {} {\bibfield  {journal} {\bibinfo  {journal} {Adv.
  Nucl. Phys.}\ }\textbf {\bibinfo {volume} {16}},\ \bibinfo {pages} {1}
  (\bibinfo {year} {1986})}\BibitemShut {NoStop}%
\bibitem [{\citenamefont {Serot}\ and\ \citenamefont
  {Walecka}(1997)}]{Serot:1997xg}%
  \BibitemOpen
  \bibfield  {author} {\bibinfo {author} {\bibfnamefont {B.~D.}\ \bibnamefont
  {Serot}}\ and\ \bibinfo {author} {\bibfnamefont {J.~D.}\ \bibnamefont
  {Walecka}},\ }\href {\doibase 10.1142/S0218301397000299} {\bibfield
  {journal} {\bibinfo  {journal} {Int. J. Mod. Phys. E}\ }\textbf {\bibinfo
  {volume} {6}},\ \bibinfo {pages} {515} (\bibinfo {year} {1997})},\ \Eprint
  {http://arxiv.org/abs/nucl-th/9701058} {arXiv:nucl-th/9701058} \BibitemShut
  {NoStop}%
\bibitem [{\citenamefont {Guichon}\ \emph {et~al.}(2018)\citenamefont
  {Guichon}, \citenamefont {Stone},\ and\ \citenamefont
  {Thomas}}]{Guichon:2018uew}%
  \BibitemOpen
  \bibfield  {author} {\bibinfo {author} {\bibfnamefont {P.}~\bibnamefont
  {Guichon}}, \bibinfo {author} {\bibfnamefont {J.}~\bibnamefont {Stone}}, \
  and\ \bibinfo {author} {\bibfnamefont {A.}~\bibnamefont {Thomas}},\ }\href
  {\doibase 10.1016/j.ppnp.2018.01.008} {\bibfield  {journal} {\bibinfo
  {journal} {Prog. Part. Nucl. Phys.}\ }\textbf {\bibinfo {volume} {100}},\
  \bibinfo {pages} {262} (\bibinfo {year} {2018})},\ \Eprint
  {http://arxiv.org/abs/1802.08368} {arXiv:1802.08368 [nucl-th]} \BibitemShut
  {NoStop}%
\bibitem [{\citenamefont {Motta}\ \emph {et~al.}(2019)\citenamefont {Motta},
  \citenamefont {Kalaitzis}, \citenamefont {Anti\'c}, \citenamefont {Guichon},
  \citenamefont {Stone},\ and\ \citenamefont {Thomas}}]{Motta:2019tjc}%
  \BibitemOpen
  \bibfield  {author} {\bibinfo {author} {\bibfnamefont {T.}~\bibnamefont
  {Motta}}, \bibinfo {author} {\bibfnamefont {A.}~\bibnamefont {Kalaitzis}},
  \bibinfo {author} {\bibfnamefont {S.}~\bibnamefont {Anti\'c}}, \bibinfo
  {author} {\bibfnamefont {P.}~\bibnamefont {Guichon}}, \bibinfo {author}
  {\bibfnamefont {J.}~\bibnamefont {Stone}}, \ and\ \bibinfo {author}
  {\bibfnamefont {A.}~\bibnamefont {Thomas}},\ }\href {\doibase
  10.3847/1538-4357/ab218e} {\bibfield  {journal} {\bibinfo  {journal}
  {Astrophys. J.}\ }\textbf {\bibinfo {volume} {878}},\ \bibinfo {pages} {159}
  (\bibinfo {year} {2019})},\ \Eprint {http://arxiv.org/abs/1904.03794}
  {arXiv:1904.03794 [nucl-th]} \BibitemShut {NoStop}%
\bibitem [{\citenamefont {Rikovska-Stone}\ \emph {et~al.}(2007)\citenamefont
  {Rikovska-Stone}, \citenamefont {Guichon}, \citenamefont {Matevosyan},\ and\
  \citenamefont {Thomas}}]{RikovskaStone:2006ta}%
  \BibitemOpen
  \bibfield  {author} {\bibinfo {author} {\bibfnamefont {J.}~\bibnamefont
  {Rikovska-Stone}}, \bibinfo {author} {\bibfnamefont {P.~A.}\ \bibnamefont
  {Guichon}}, \bibinfo {author} {\bibfnamefont {H.~H.}\ \bibnamefont
  {Matevosyan}}, \ and\ \bibinfo {author} {\bibfnamefont {A.~W.}\ \bibnamefont
  {Thomas}},\ }\href {\doibase 10.1016/j.nuclphysa.2007.05.011} {\bibfield
  {journal} {\bibinfo  {journal} {Nucl. Phys. A}\ }\textbf {\bibinfo {volume}
  {792}},\ \bibinfo {pages} {341} (\bibinfo {year} {2007})},\ \Eprint
  {http://arxiv.org/abs/nucl-th/0611030} {arXiv:nucl-th/0611030} \BibitemShut
  {NoStop}%
\bibitem [{\citenamefont {Guichon}\ and\ \citenamefont
  {Thomas}(2004)}]{Guichon:2004xg}%
  \BibitemOpen
  \bibfield  {author} {\bibinfo {author} {\bibfnamefont {P.~A.}\ \bibnamefont
  {Guichon}}\ and\ \bibinfo {author} {\bibfnamefont {A.~W.}\ \bibnamefont
  {Thomas}},\ }\href {\doibase 10.1103/PhysRevLett.93.132502} {\bibfield
  {journal} {\bibinfo  {journal} {Phys. Rev. Lett.}\ }\textbf {\bibinfo
  {volume} {93}},\ \bibinfo {pages} {132502} (\bibinfo {year} {2004})},\
  \Eprint {http://arxiv.org/abs/nucl-th/0402064} {arXiv:nucl-th/0402064}
  \BibitemShut {NoStop}%
\bibitem [{\citenamefont {Reddy}\ \emph {et~al.}(1998)\citenamefont {Reddy},
  \citenamefont {Prakash},\ and\ \citenamefont {Lattimer}}]{Reddy:1997yr}%
  \BibitemOpen
  \bibfield  {author} {\bibinfo {author} {\bibfnamefont {S.}~\bibnamefont
  {Reddy}}, \bibinfo {author} {\bibfnamefont {M.}~\bibnamefont {Prakash}}, \
  and\ \bibinfo {author} {\bibfnamefont {J.~M.}\ \bibnamefont {Lattimer}},\
  }\href {\doibase 10.1103/PhysRevD.58.013009} {\bibfield  {journal} {\bibinfo
  {journal} {Phys. Rev.}\ }\textbf {\bibinfo {volume} {D58}},\ \bibinfo {pages}
  {013009} (\bibinfo {year} {1998})},\ \Eprint
  {http://arxiv.org/abs/astro-ph/9710115} {arXiv:astro-ph/9710115 [astro-ph]}
  \BibitemShut {NoStop}%
\bibitem [{\citenamefont {Ozel}\ and\ \citenamefont
  {Freire}(2016)}]{Ozel:2016oaf}%
  \BibitemOpen
  \bibfield  {author} {\bibinfo {author} {\bibfnamefont {F.}~\bibnamefont
  {Ozel}}\ and\ \bibinfo {author} {\bibfnamefont {P.}~\bibnamefont {Freire}},\
  }\href {\doibase 10.1146/annurev-astro-081915-023322} {\bibfield  {journal}
  {\bibinfo  {journal} {Ann. Rev. Astron. Astrophys.}\ }\textbf {\bibinfo
  {volume} {54}},\ \bibinfo {pages} {401} (\bibinfo {year} {2016})},\ \Eprint
  {http://arxiv.org/abs/1603.02698} {arXiv:1603.02698 [astro-ph.HE]}
  \BibitemShut {NoStop}%
\bibitem [{\citenamefont {Antoniadis}\ \emph {et~al.}(2016)\citenamefont
  {Antoniadis}, \citenamefont {Tauris}, \citenamefont {Ozel}, \citenamefont
  {Barr}, \citenamefont {Champion},\ and\ \citenamefont
  {Freire}}]{Antoniadis:2016hxz}%
  \BibitemOpen
  \bibfield  {author} {\bibinfo {author} {\bibfnamefont {J.}~\bibnamefont
  {Antoniadis}}, \bibinfo {author} {\bibfnamefont {T.~M.}\ \bibnamefont
  {Tauris}}, \bibinfo {author} {\bibfnamefont {F.}~\bibnamefont {Ozel}},
  \bibinfo {author} {\bibfnamefont {E.}~\bibnamefont {Barr}}, \bibinfo {author}
  {\bibfnamefont {D.~J.}\ \bibnamefont {Champion}}, \ and\ \bibinfo {author}
  {\bibfnamefont {P.~C.~C.}\ \bibnamefont {Freire}},\ }\href@noop {} {\
  (\bibinfo {year} {2016})},\ \Eprint {http://arxiv.org/abs/1605.01665}
  {arXiv:1605.01665 [astro-ph.HE]} \BibitemShut {NoStop}%
\bibitem [{\citenamefont {Cirelli}\ \emph {et~al.}(2013)\citenamefont
  {Cirelli}, \citenamefont {Del~Nobile},\ and\ \citenamefont
  {Panci}}]{DelNobile:2013sia}%
  \BibitemOpen
  \bibfield  {author} {\bibinfo {author} {\bibfnamefont {M.}~\bibnamefont
  {Cirelli}}, \bibinfo {author} {\bibfnamefont {E.}~\bibnamefont {Del~Nobile}},
  \ and\ \bibinfo {author} {\bibfnamefont {P.}~\bibnamefont {Panci}},\ }\href
  {\doibase 10.1088/1475-7516/2013/10/019} {\bibfield  {journal} {\bibinfo
  {journal} {JCAP}\ }\textbf {\bibinfo {volume} {1310}},\ \bibinfo {pages}
  {019} (\bibinfo {year} {2013})},\ \Eprint {http://arxiv.org/abs/1307.5955}
  {arXiv:1307.5955 [hep-ph]} \BibitemShut {NoStop}%
\bibitem [{\citenamefont {Thomas}\ and\ \citenamefont
  {Weise}(2001)}]{Thomas:2001kw}%
  \BibitemOpen
  \bibfield  {author} {\bibinfo {author} {\bibfnamefont {A.~W.}\ \bibnamefont
  {Thomas}}\ and\ \bibinfo {author} {\bibfnamefont {W.}~\bibnamefont {Weise}},\
  }\href {\doibase 10.1002/352760314X} {\emph {\bibinfo {title} {{The Structure
  of the Nucleon}}}}\ (\bibinfo  {publisher} {Wiley},\ \bibinfo {address}
  {Germany},\ \bibinfo {year} {2001})\BibitemShut {NoStop}%
\bibitem [{\citenamefont {Zanotti}\ \emph {et~al.}(2017)\citenamefont
  {Zanotti}, \citenamefont {Bickerton}, \citenamefont {Horsley}, \citenamefont
  {Nakamura}, \citenamefont {Rakow}, \citenamefont {Schierholz}, \citenamefont
  {Shanahan},\ and\ \citenamefont {Young}}]{Zanotti:2017bte}%
  \BibitemOpen
  \bibfield  {author} {\bibinfo {author} {\bibfnamefont {J.}~\bibnamefont
  {Zanotti}}, \bibinfo {author} {\bibfnamefont {J.}~\bibnamefont {Bickerton}},
  \bibinfo {author} {\bibfnamefont {R.}~\bibnamefont {Horsley}}, \bibinfo
  {author} {\bibfnamefont {Y.}~\bibnamefont {Nakamura}}, \bibinfo {author}
  {\bibfnamefont {P.}~\bibnamefont {Rakow}}, \bibinfo {author} {\bibfnamefont
  {G.}~\bibnamefont {Schierholz}}, \bibinfo {author} {\bibfnamefont
  {P.}~\bibnamefont {Shanahan}}, \ and\ \bibinfo {author} {\bibfnamefont
  {R.}~\bibnamefont {Young}} (\bibinfo {collaboration} {QCDSF/UKQCD}),\ }\href
  {\doibase 10.22323/1.256.0163} {\bibfield  {journal} {\bibinfo  {journal}
  {PoS}\ }\textbf {\bibinfo {volume} {LATTICE2016}},\ \bibinfo {pages} {163}
  (\bibinfo {year} {2017})}\BibitemShut {NoStop}%
\bibitem [{\citenamefont {Alarc\'on}\ and\ \citenamefont
  {Weiss}(2017)}]{Alarcon:2017ivh}%
  \BibitemOpen
  \bibfield  {author} {\bibinfo {author} {\bibfnamefont {J.~M.}\ \bibnamefont
  {Alarc\'on}}\ and\ \bibinfo {author} {\bibfnamefont {C.}~\bibnamefont
  {Weiss}},\ }\href {\doibase 10.1103/PhysRevC.96.055206} {\bibfield  {journal}
  {\bibinfo  {journal} {Phys. Rev. C}\ }\textbf {\bibinfo {volume} {96}},\
  \bibinfo {pages} {055206} (\bibinfo {year} {2017})},\ \Eprint
  {http://arxiv.org/abs/1707.07682} {arXiv:1707.07682 [hep-ph]} \BibitemShut
  {NoStop}%
\bibitem [{\citenamefont {{Goriely}}\ \emph {et~al.}(2013)\citenamefont
  {{Goriely}}, \citenamefont {{Chamel}},\ and\ \citenamefont
  {{Pearson}}}]{Goriely:2013}%
  \BibitemOpen
  \bibfield  {author} {\bibinfo {author} {\bibfnamefont {S.}~\bibnamefont
  {{Goriely}}}, \bibinfo {author} {\bibfnamefont {N.}~\bibnamefont {{Chamel}}},
  \ and\ \bibinfo {author} {\bibfnamefont {J.~M.}\ \bibnamefont {{Pearson}}},\
  }\href {\doibase 10.1103/PhysRevC.88.024308} {\bibfield  {journal} {\bibinfo
  {journal} {Phys. Rev. C}\ }\textbf {\bibinfo {volume} {88}},\ \bibinfo {eid}
  {024308} (\bibinfo {year} {2013})}\BibitemShut {NoStop}%
\bibitem [{\citenamefont {Pearson}\ \emph {et~al.}(2018)\citenamefont
  {Pearson}, \citenamefont {Chamel}, \citenamefont {Potekhin}, \citenamefont
  {Fantina}, \citenamefont {Ducoin}, \citenamefont {Dutta},\ and\ \citenamefont
  {Goriely}}]{Pearson:2018tkr}%
  \BibitemOpen
  \bibfield  {author} {\bibinfo {author} {\bibfnamefont {J.~M.}\ \bibnamefont
  {Pearson}}, \bibinfo {author} {\bibfnamefont {N.}~\bibnamefont {Chamel}},
  \bibinfo {author} {\bibfnamefont {A.~Y.}\ \bibnamefont {Potekhin}}, \bibinfo
  {author} {\bibfnamefont {A.~F.}\ \bibnamefont {Fantina}}, \bibinfo {author}
  {\bibfnamefont {C.}~\bibnamefont {Ducoin}}, \bibinfo {author} {\bibfnamefont
  {A.~K.}\ \bibnamefont {Dutta}}, \ and\ \bibinfo {author} {\bibfnamefont
  {S.}~\bibnamefont {Goriely}},\ }\href {\doibase 10.1093/mnras/sty2413,
  10.1093/mnras/stz800} {\bibfield  {journal} {\bibinfo  {journal} {Mon. Not.
  Roy. Astron. Soc.}\ }\textbf {\bibinfo {volume} {481}},\ \bibinfo {pages}
  {2994} (\bibinfo {year} {2018})},\ \bibinfo {note} {[erratum: Mon. Not. Roy.
  Astron. Soc.486,no.1,768(2019)]},\ \Eprint {http://arxiv.org/abs/1903.04981}
  {arXiv:1903.04981 [astro-ph.HE]} \BibitemShut {NoStop}%
\bibitem [{\citenamefont {Chamel}\ \emph {et~al.}(2009)\citenamefont {Chamel},
  \citenamefont {Goriely},\ and\ \citenamefont {Pearson}}]{Chamel:2009yx}%
  \BibitemOpen
  \bibfield  {author} {\bibinfo {author} {\bibfnamefont {N.}~\bibnamefont
  {Chamel}}, \bibinfo {author} {\bibfnamefont {S.}~\bibnamefont {Goriely}}, \
  and\ \bibinfo {author} {\bibfnamefont {J.}~\bibnamefont {Pearson}},\ }\href
  {\doibase 10.1103/PhysRevC.80.065804} {\bibfield  {journal} {\bibinfo
  {journal} {Phys. Rev. C}\ }\textbf {\bibinfo {volume} {80}},\ \bibinfo
  {pages} {065804} (\bibinfo {year} {2009})},\ \Eprint
  {http://arxiv.org/abs/0911.3346} {arXiv:0911.3346 [nucl-th]} \BibitemShut
  {NoStop}%
\end{thebibliography}%

    \foreach \x in {1,...,\numbersupplementpages}
    {
        \clearpage
        \includepdf[pages={\x,{}}]{\supplementfilename}
    }

\end{document}